\documentclass[11pt,a4paper]{article}

\usepackage[T1]{fontenc}
\usepackage[utf8]{inputenc}
\usepackage{lmodern}
\usepackage{amsmath,amssymb,amsthm}
\usepackage{booktabs,graphicx,placeins}
\usepackage{needspace}
\usepackage[margin=1in]{geometry}
\usepackage{microtype}
\usepackage[numbers,sort]{natbib}
\usepackage[hidelinks]{hyperref}
\hypersetup{
  pdftitle={Accelerating Computation and Resolving Extraneous Roots in the Kinetic Plasma Dispersion Relation Solver BO},
  pdfauthor={Xizhi Tian; Huasheng Xie; Long Wei},
  pdfsubject={Kinetic plasma dispersion relations},
  pdfkeywords={Kinetic dispersion relation, Plasma waves and instabilities, Extraneous roots, Matrix reduction, Root screening, BO/PDRK solver}
}

\newtheorem{theorem}{Theorem}[section]
\newtheorem{proposition}[theorem]{Proposition}
\newtheorem{lemma}[theorem]{Lemma}
\newtheorem{corollary}[theorem]{Corollary}
\AddToHook{env/theorem/before}{\Needspace{6\baselineskip}}
\AddToHook{env/proposition/before}{\Needspace{6\baselineskip}}
\AddToHook{env/lemma/before}{\Needspace{6\baselineskip}}
\AddToHook{env/corollary/before}{\Needspace{6\baselineskip}}
\theoremstyle{remark}
\newtheorem{remark}[theorem]{Remark}
\AddToHook{env/thebibliography/begin}{\interlinepenalty=10000}

\title{\textbf{Accelerating Computation and Resolving Extraneous Roots in the Kinetic Plasma Dispersion Relation Solver BO}}

\author{
  Xizhi Tian$^{1}$, Huasheng Xie$^{2,*}$, Long Wei$^{1,3,*}$ \\[0.6em]
  \parbox{0.97\textwidth}{\centering
  \small $^{1}$Fudan University, Shanghai 200433, People's Republic of China \\
  \small $^{2}$Beijing VeloAlpha Technology Co. Ltd, Beijing 100080, People's Republic of China \\
  \small $^{3}$Shanghai Academy of AI for Science, Shanghai 200232, People's Republic of China \\[0.3em]
  \footnotesize $^{*}$Corresponding authors.\\
  \footnotesize E-mail addresses: \texttt{tianxizhi@fudan.edu.cn} (X. Tian),\\
  \footnotesize \texttt{huashengxie@gmail.com} (H. Xie), \texttt{weilong@fudan.edu.cn} (L. Wei)
  }
}
\date{}

\begin{document}

\maketitle
\vspace{-1.5em}

\begin{abstract}
\noindent
Kinetic dispersion relations are widely used to study waves and instabilities in space, astrophysical, and laboratory plasmas. Iterative root searches depend on initial guesses and may fail to converge or miss important branches. The BO/PDRK solvers [Comput.\ Phys.\ Commun.\ 244, 343--371 (2019)] use rational pole expansions to transform the nonlinear dispersion relation into a matrix eigenvalue problem, obtaining all roots of the approximated relation in one calculation. However, high cyclotron harmonics enlarge the matrix and slow computation, while extraneous roots are difficult to separate from physical modes. We develop BO-T (BO-Tian), an improved version of BO combining two methods. First, for matrix reduction, BO-T applies rank factorization to combined residue matrices, preserving the rational dispersion relation and accounting for the removed pole eigenvalues. Second, for root screening, BO-T uses contour comparison between approximate and reference kinetic dispersion matrices along a closed contour enclosing no poles. This provides a sufficient condition for preserving local root counts, with error bounds for isolated roots and criteria for growth rate signs. Benchmarks on representative electrostatic and electromagnetic cases show speedups of about 10 to 140 for matrix assembly and eigenvalue calculation, exceeding 30 in most cases with high cyclotron harmonics. Root screening recovers the main unstable branches in agreement with published and archived dispersion curves.

\end{abstract}

\vspace{0.5em}

\section{Introduction}
Across the solar wind, Earth's magnetosphere, astrophysical environments, and laboratory plasmas, small disturbances propagate as collective waves, decay through kinetic damping, or grow into instabilities \cite{astfalkJenko2017leopard,verscharenEtAl2018alps}. Linear kinetic theory describes these behaviors through dispersion relations derived from the linearized Vlasov--Maxwell system \cite{verscharenEtAl2018alps,xieXiao2016pdrk}, where the complex roots determine the real frequency and growth or damping rate of each mode. Except in simplified cases, realistic multi-component magnetized plasmas support many coexisting wave branches. Conventional iterative root-finding methods are often slow, sensitive to the initial guess, and may fail to converge, so physically important branches can be missed \cite{xieXiao2016pdrk}.

The BO/PDRK family, originating with PDRK, provides an efficient alternative to iterative root searches \cite{xieXiao2016pdrk,xie2019bo}. By approximating the plasma dispersion function with a $J$-pole rational expansion, these solvers transform the nonlinear dispersion relation into an equivalent matrix eigenvalue problem. A single eigenvalue calculation then gives all roots of the approximated relation without requiring an initial guess for each mode. Subsequent versions have extended this framework to a wide range of plasma models and equilibrium velocity distributions, including kappa-type and arbitrary gyrotropic distributions \cite{xieEtAl2021bo2,baiEtAl2025boKM,xie2025boarbitrary,baiXie2026boPBK}.

However, the BO/PDRK solvers still face two practical problems. First, the matrix eigenvalue calculation becomes computationally expensive when the matrix dimension is large. Increasing the number of plasma species, retained cyclotron harmonics, or rational poles rapidly enlarges the augmented matrix, and higher-order expansions of the velocity distribution can also require higher pole orders \cite{xieXiao2016pdrk,xie2019bo,xie2025boarbitrary,baiXie2026boPBK}. This makes full-spectrum calculations over multi-dimensional wave-vector or parameter scans very time-consuming, whereas using lower truncation orders to reduce the matrix size can leave the dispersion branches unconverged \cite{xieXiao2016pdrk,xie2019bo,xie2025boarbitrary}. Second, the computed spectrum contains numerous extraneous roots that are difficult to separate from physical modes. Extraneous roots from the rational approximation can be less damped than the physical modes, and approximation errors can even produce artificial positive growth rates \cite{xieXiao2016pdrk,xie2019bo}. In addition, errors in fitting the velocity distribution can shift the growth rates and unstable wave-number ranges, or even miss genuine unstable branches and introduce spurious ones \cite{guoEtAl2026alpsbo}. Consequently, physical modes cannot be identified simply by sorting the computed eigenvalues by their growth or damping rates.

To address these two problems, we use rank factorization of the combined residue matrices for matrix reduction and contour comparison of the dispersion matrices for root screening. Specifically, to reduce the matrix dimension, we exploit the fact that many pole terms in the BO/PDRK formulation are repeated across species, harmonics, and basis functions. By grouping identical poles and applying singular value decomposition (SVD) to their combined residue matrices, redundant auxiliary variables can be eliminated without changing the rational dispersion relation \cite{suBai2011rational,amparanEtAl2018rational}. For root screening, we evaluate the difference between the approximate dispersion matrix and the reference kinetic dispersion matrix along a closed contour enclosing a candidate root and no poles \cite{bindelHood2013localization}. We compare this boundary difference with the minimum singular value of the approximate dispersion matrix along the contour. The resulting criterion gives a sufficient condition for preserving the local root count and bounding the errors in frequency and growth rate for an isolated root.

In this work, we incorporate both methods into an improved version of BO, named BO-T (BO-Tian). On the theoretical side, we derive the exact determinant identity and eigenvalue multiplicity formula connecting the full and reduced matrices, and establish a contour perturbation criterion that gives the local root count, the error radius for isolated roots, and the sign of their growth rates. On the numerical side, we implement these procedures for the electrostatic, electromagnetic, and Hermite--Hermite solvers in BO and evaluate them across thirteen benchmark cases. The complete BO/PDRK spectrum is reconstructed from the reduced eigenvalues and the removed pole eigenvalues. Timing tests on eleven cases show speedups of about 10 to 140 for matrix assembly and eigenvalue calculation, with speedups above 30 in most cases with high cyclotron harmonics. Moreover, root screening by contour comparison automatically recovers the main unstable branches in good agreement with published and archived dispersion curves, greatly improving the efficiency and usability of the BO framework for kinetic plasma wave and instability analysis.

\section{Problem setup}
\label{sec:problem-setup}

\subsection{Linear kinetic dispersion relation}
\label{sec:kinetic-response}

We consider a uniform plasma with background magnetic field $\mathbf B_0=(0,0,B_0)$. Each species $s=1,\ldots,S$ has equilibrium distribution function $f_{s0}$, charge $q_s$, mass $m_s$, and density $n_{s0}$. We take the wave vector to be $\mathbf k=(k_\perp,0,k_\parallel)=(k\sin\theta,0,k\cos\theta)$ \cite{xieXiao2016pdrk,xie2019bo,xieEtAl2021bo2,xie2025boarbitrary}.

Following the convention in PDRK and BO, we write linear plane-wave perturbations as
\begin{equation}
  \delta\psi(\mathbf x,t)
  =\widehat{\psi}\exp\!\left[\mathrm{i}(\mathbf k\!\cdot\!\mathbf x-\omega t)\right],
  \qquad
  \omega=\omega_r+\mathrm{i}\gamma,
  \qquad
  \gamma=\operatorname{Im}\omega ,
  \label{eq:plane-wave-convention}
\end{equation}
where $\omega_r$ is the real frequency and $\gamma$ is the growth rate. With this convention, $\gamma>0$ indicates instability and $\gamma<0$ indicates damping.

From the linearized Vlasov and Maxwell equations, the dispersion relation can be written as
\begin{equation}
  \mathbf D_\star(\omega,\mathbf k;\mathcal F_0)\widehat{\mathbf E}=\mathbf 0,
  \qquad
  \widehat{\mathbf E}\ne\mathbf 0,
  \qquad
  \mathcal F_0=\{f_{s0}\}_{s=1}^{S}.
  \label{eq:target-response}
\end{equation}
Here $\mathbf D_\star$ is exact within the chosen kinetic model, with its specified equilibrium distributions and field equations, before rational approximation and numerical truncation. In the electrostatic (ES) limit, $\mathbf D_\star$ is a scalar function; in the full electromagnetic (EM) and Darwin models, it is a $3\times3$ matrix. For the full electromagnetic case,
\begin{equation}
  \mathbf D_\star(\omega,\mathbf k)
  =\mathbf K_\star(\omega,\mathbf k)
  +\frac{c^2}{\omega^2}
  \left(\mathbf k\mathbf k^{\mathsf T}-k^2\mathbf I\right),
  \label{eq:electromagnetic-response}
\end{equation}
where $\mathbf K_\star$ is the dielectric tensor \cite{xieXiao2016pdrk,xie2019bo}.

For a fixed wave vector $\mathbf k$, let $\Omega\subset\mathbb C$ be an open domain on which $\mathbf D_\star$ is analytic and $\det\mathbf D_\star$ does not vanish identically. The target dispersion function and its roots on $\Omega$ are
\begin{equation}
  d_\star(\omega,\mathbf k)
  :=\det\mathbf D_\star(\omega,\mathbf k),
  \qquad
  \Lambda(\mathbf D_\star;\Omega)
  :=\left\{\omega\in\Omega:d_\star(\omega,\mathbf k)=0\right\}.
  \label{eq:target-spectrum}
\end{equation}
The algebraic multiplicity of a root in $\Lambda(\mathbf D_\star;\Omega)$ is its order as a zero of $d_\star(\omega,\mathbf k)$ \cite{guttelTisseur2017nep}. Throughout this paper, the subscript $\star$ refers to this fixed kinetic model.

\subsection{Rational approximation in BO/PDRK}
\label{sec:finite-response}

To transform the kinetic dispersion relation into an algebraic eigenvalue problem, PDRK and BO approximate the plasma dispersion function $Z(\zeta)$ by a $J$-pole expansion \cite{xieXiao2016pdrk},
\begin{equation}
  Z(\zeta)\approx Z_J(\zeta)
  =\sum_{j=1}^{J}\frac{b_j}{\zeta-c_j}.
  \label{eq:j-pole-expansion}
\end{equation}
In magnetized plasmas, the summation over Bessel functions and cyclotron harmonics is truncated to $n=-N,\ldots,N$. For basis-expansion solvers such as BO-Arbitrary, the velocity-space expansion is also truncated at $(l_{\max},m_{\max})$ \cite{xie2025boarbitrary}. We collect the approximation and truncation parameters in $h=(J,N,l_{\max},m_{\max},\ldots)$. For a fixed $h$, $\mathbf D_h$ denotes the resulting approximate dispersion matrix in exact arithmetic.

With these approximations, $\mathbf D_h(\omega,\mathbf k)$ is a rational matrix function of $\omega$ and can be written in partial-fraction form as
\begin{equation}
  \mathbf D_h(\omega,\mathbf k)
  =\mathbf P_h(\omega,\mathbf k)
  +\sum_{g=1}^{G_h}\sum_{\ell=1}^{\nu_g}
  \frac{\mathbf R^{(h)}_{g,\ell}(\mathbf k)}
       {(\omega-\xi_{g,h}(\mathbf k))^{\ell}},
  \label{eq:finite-rational-response}
\end{equation}
where $\mathbf P_h$ is a matrix polynomial in $\omega$, $G_h$ is the number of distinct poles $\xi_{g,h}(\mathbf k)$, $\nu_g$ is the order of the $g$th pole, and $\mathbf R^{(h)}_{g,\ell}(\mathbf k)$ are the corresponding coefficient matrices. We denote the pole set of $\mathbf D_h$ by $\mathcal P(\mathbf D_h)$.

The difference between the target and approximate dispersion matrices is
\begin{equation}
  \Delta\mathbf D_h(\omega,\mathbf k)
  :=\mathbf D_\star(\omega,\mathbf k)-\mathbf D_h(\omega,\mathbf k).
  \label{eq:response-discrepancy}
\end{equation}
This difference contains the errors from fitting the equilibrium distribution, the $J$-pole expansion, and harmonic truncation.

For two choices $h_1$ and $h_2$ of $h$, $\mathbf D_{h_1}$ and $\mathbf D_{h_2}$ are the matrices in Eq.~\eqref{eq:finite-rational-response} with $h$ replaced by $h_1$ and $h_2$, respectively. Their difference measures agreement between the two finite approximations. The error relative to the target matrix $\mathbf D_\star$ in Eq.~\eqref{eq:target-response} is instead $\Delta\mathbf D_h$ in Eq.~\eqref{eq:response-discrepancy}. All comparisons use the same field variables, frequency normalization, and matrix scaling.

\subsection{Matrix eigenvalue formulation}
\label{sec:augmented-realization}

By introducing auxiliary variables for the rational terms in Eq.~\eqref{eq:finite-rational-response}, PDRK and BO transform the approximate dispersion relation into a linear generalized eigenvalue problem,
\begin{equation}
  \mathcal L_h^{\mathrm{full}}(\omega,\mathbf k)\mathbf x_h
  :=\left(\mathbf A_h^{\mathrm{full}}(\mathbf k)
  -\omega\mathbf B_h^{\mathrm{full}}(\mathbf k)\right)\mathbf x_h
  =\mathbf 0,
  \label{eq:full-pencil}
\end{equation}
which reduces to the standard eigenvalue problem $\mathbf A_h^{\mathrm{full}}\mathbf x_h=\omega\mathbf x_h$ when $\mathbf B_h^{\mathrm{full}}=\mathbf I$ \cite{xieXiao2016pdrk,xie2019bo,xieEtAl2021bo2}. The vector $\mathbf x_h$ contains the auxiliary variables for the rational terms and the retained field and current components, including the electric and magnetic fields in EM solvers. At frequencies where the required eliminations are nonsingular, eliminating the auxiliary variables and other retained components recovers $\mathbf D_h(\omega,\mathbf k)\widehat{\mathbf E}=\mathbf 0$ \cite{suBai2011rational,amparanEtAl2018rational}. We assume that the matrix pair is regular, meaning that $\det\mathcal L_h^{\mathrm{full}}(\omega,\mathbf k)$ does not vanish identically.

The roots of the approximate dispersion relation on $\Omega$ are
\begin{equation}
  \Lambda(\mathbf D_h;\Omega)
  :=\left\{\omega\in\Omega\setminus\mathcal P(\mathbf D_h):
  \det\mathbf D_h(\omega,\mathbf k)=0\right\}.
  \label{eq:finite-response-spectrum}
\end{equation}
We assume that $\det\mathbf D_h$ does not vanish identically on the regions under consideration and count each root by its order as a zero of this determinant. Thus there are three distinct spectra: the target roots $\Lambda(\mathbf D_\star;\Omega)$, the approximate roots $\Lambda(\mathbf D_h;\Omega)$, and the eigenvalues of the matrix pair $(\mathbf A_h^{\mathrm{full}},\mathbf B_h^{\mathrm{full}})$. Two sources of extra solutions are relevant here:

\begin{enumerate}
  \item Separate species, harmonics, or basis terms can contribute to the same pole. If their combined contribution is represented with more auxiliary variables than it requires, the excess variables can produce additional eigenvalues at that pole.
  \item The rational approximation and finite truncations can introduce roots of $\det\mathbf D_h=0$ that do not approximate roots of the target relation. These are roots of the approximate dispersion relation itself, away from its poles, even when no nearby target root exists.
\end{enumerate}

\section{Matrix reduction and root screening}
\label{sec:methodology}

BO-T first reduces the matrix eigenvalue problem without changing the finite rational dispersion relation. It then compares the resulting dispersion matrix with a reference matrix to assess the computed roots.

\subsection{Matrix reduction}
\label{sec:response-preserving-reduction}

We first collect auxiliary variables that share the same pole. For the simple pole blocks considered here, this grouping gives the regular linear matrix function
\begin{equation}
  \mathcal L_h^{\mathrm{full}}(\omega)
  =
  \begin{bmatrix}
    \boldsymbol\Xi-\omega\mathbf I_b & \mathbf B\\
    \mathbf C & \mathbf F-\omega\mathbf M
  \end{bmatrix},
  \qquad
  \boldsymbol\Xi
  =\operatorname{diag}
  \left(\xi_1\mathbf I_{b_1},\ldots,\xi_G\mathbf I_{b_G}\right).
  \label{eq:full-grouped-pencil}
\end{equation}
Here $\xi_g$ is the pole of group $g$, $b_g$ is the number of auxiliary variables in that group, and $b=\sum_g b_g$. The $m\times m$ block $\mathbf F-\omega\mathbf M$ contains the retained field and current variables. In electromagnetic models these include the electric field, magnetic field, and retained current components; $m$ depends on the formulation. All pole locations are kept unchanged, so only exactly equal poles are grouped.

To see which auxiliary variables are needed, eliminate the upper block for $\omega\ne\xi_g$. Partitioning the coupling matrices into $\mathbf B_g\in\mathbb C^{b_g\times m}$ and $\mathbf C_g\in\mathbb C^{m\times b_g}$ gives the Schur complement
\begin{equation}
  \mathbf S_h(\omega)
  =\mathbf F-\omega\mathbf M
   +\sum_{g=1}^{G}
      \frac{\mathbf C_g\mathbf B_g}{\omega-\xi_g}.
  \label{eq:preserved-schur-response}
\end{equation}
Thus the retained equations depend on the $b_g$ variables in group $g$ only through the combined residue matrix
\begin{equation}
  \mathbf Q_g=\mathbf C_g\mathbf B_g,
  \qquad
  r_g=\operatorname{rank}\mathbf Q_g\le\min(b_g,m).
  \label{eq:grouped-residue}
\end{equation}
The rank $r_g$ limits the number of auxiliary variables required for this pole, regardless of how many terms were assembled into the group. In the full electromagnetic kernels implemented here, the pole variables are driven by the three electric field components, and their response enters the electric field equations through an embedded $3\times3$ residue. These residue matrices therefore have rank at most three, even when more field and current variables are retained. The corresponding electrostatic residue is scalar and has rank at most one. Combining terms at an identical pole can therefore replace several auxiliary blocks by a single block of this bounded size; further cancellation can lower its rank.

When $r_g<b_g$, the same residue can be represented with fewer auxiliary variables. Choose an exact rank factorization $\mathbf Q_g=\mathbf L_g\mathbf R_g$, where $\mathbf L_g\in\mathbb C^{m\times r_g}$ and $\mathbf R_g\in\mathbb C^{r_g\times m}$, and define
\begin{equation}
  \boldsymbol\Xi_r
  =\operatorname{diag}
  \left(\xi_1\mathbf I_{r_1},\ldots,\xi_G\mathbf I_{r_G}\right),
  \quad
  \mathbf R=\operatorname{col}(\mathbf R_1,\ldots,\mathbf R_G),
  \quad
  \mathbf L=[\mathbf L_1,\ldots,\mathbf L_G].
  \label{eq:reduced-factors}
\end{equation}
The reduced eigenvalue problem is then defined by
\begin{equation}
  \mathcal L_h^{\mathrm{redu}}(\omega)
  =
  \begin{bmatrix}
    \boldsymbol\Xi_r-\omega\mathbf I_r & \mathbf R\\
    \mathbf L & \mathbf F-\omega\mathbf M
  \end{bmatrix},
  \qquad r=\sum_g r_g.
  \label{eq:reduced-grouped-pencil}
\end{equation}
The full matrix has dimension $b+m$, whereas the reduced matrix has dimension $r+m$. Eliminating either auxiliary block gives the same $\mathbf S_h$, because $\mathbf L_g\mathbf R_g=\mathbf C_g\mathbf B_g$. The residue ranks also give the smallest auxiliary dimension for this simple pole sum, as proved in Proposition~\ref{prop:app-minimality} using the theory of rational realizations \cite{suBai2011rational,amparanEtAl2018rational,dopicoEtAl2020local}.

The relation between the full and reduced spectra is expressed by
\begin{equation}
  \det\mathcal L_h^{\mathrm{full}}(\omega)
  =\det\mathcal L_h^{\mathrm{redu}}(\omega)
   \prod_{g=1}^{G}(\xi_g-\omega)^{b_g-r_g}.
  \label{eq:determinant-accounting}
\end{equation}
This identity holds for all $\omega\in\mathbb C$ in exact arithmetic (Theorem~\ref{thm:exact-reduction}). Every eigenvalue away from the poles retains its algebraic multiplicity, and $b_g-r_g$ eigenvalues at each $\xi_g$ are removed. The complete finite spectrum of the original matrix is therefore recovered by appending these known pole eigenvalues to the reduced spectrum.

The unchanged Schur complement also gives the same $\mathbf D_h$ after identical nonsingular eliminations and scalings of the remaining variables. In the electromagnetic case, this step eliminates the magnetic field and retained current components to obtain the electric field dispersion matrix. The regularity conditions are stated in Eq.~\eqref{eq:app-regular-elimination}. The auxiliary groups counted by $G$ need not coincide with the actual poles of $\mathbf D_h$ counted by $G_h$ in Eq.~\eqref{eq:finite-rational-response}: some auxiliary pole terms cancel, and later field eliminations can introduce zero frequency factors.

In floating-point arithmetic, SVD and a numerical threshold give a rank $k_g$ and approximate factors $\widetilde{\mathbf Q}_g=\widetilde{\mathbf L}_g\widetilde{\mathbf R}_g$. The rank $k_g$ may differ from $r_g$. Replacing $\mathbf Q_g$ by $\widetilde{\mathbf Q}_g$ defines the approximate Schur complement $\widetilde{\mathbf S}_h$, whose change satisfies
\begin{equation}
  \left\|\widetilde{\mathbf S}_h(\omega)-\mathbf S_h(\omega)\right\|_2
  \le
  \sum_{g=1}^{G}
  \frac{\left\|\widetilde{\mathbf Q}_g-\mathbf Q_g\right\|_2}
       {|\omega-\xi_g|},
  \qquad \omega\notin\{\xi_1,\ldots,\xi_G\}.
  \label{eq:reduction-response-defect}
\end{equation}
This bound connects the numerical rank decision to the accuracy of the dispersion calculation. For an exact SVD, the residue error is the largest discarded singular value; the computed factors also carry roundoff and residue assembly errors. The denominators show why the same residue error matters more near a pole. The bound first applies to $\mathbf S_h$; Appendix~\ref{app:truncation} gives its proof and the propagation through subsequent eliminations to $\mathbf D_h$. We record the singular values to inspect the gap at the chosen threshold.

\subsection{Root screening by contour comparison}
\label{sec:root-localization}

Root screening compares two small dispersion matrices: $\mathbf D_0$, used to generate the candidate roots, and $\mathbf D_1$, chosen as the reference. Both use the common variables, normalization, and scaling specified in Section~\ref{sec:finite-response}. With exact residue factorization, $\mathbf D_0=\mathbf D_h$ from Eq.~\eqref{eq:finite-rational-response}. If the SVD is truncated numerically, $\mathbf D_0$ is obtained from $\widetilde{\mathbf S}_h$ by the same regular field eliminations.

The reference $\mathbf D_1$ is either $\mathbf D_{h_1}$, constructed from Eq.~\eqref{eq:finite-rational-response} at a comparison configuration $h_1$, or $\mathbf D_\star$ from Eq.~\eqref{eq:target-response}. The first choice assesses agreement between two finite approximations. The second assesses error relative to the specified kinetic model.

A small residual at a candidate frequency measures how nearly the equation is satisfied at that point. It does not, by itself, bound the distance to a root of the reference matrix. We instead compare the matrices on a contour surrounding the candidate, so that the comparison controls the roots throughout the enclosed region.

Let $\Gamma=\partial U$ be a positively oriented, piecewise smooth simple closed contour. Both matrices are assumed analytic on a neighborhood of $\overline U$, and $\mathbf D_0$ is nonsingular on $\Gamma$. Thus neither the interior nor the boundary contains a pole of either matrix. Define the matrix difference and its relative size on the boundary by
\begin{equation}
  \mathbf E(\omega)=\mathbf D_1(\omega)-\mathbf D_0(\omega),
  \qquad
  q_\Gamma=\sup_{\omega\in\Gamma}
    \left\|\mathbf D_0(\omega)^{-1}\mathbf E(\omega)\right\|_2.
  \label{eq:relative-boundary-discrepancy}
\end{equation}
The comparison follows the continuous change from $\mathbf D_0$ to $\mathbf D_1$,
\begin{equation}
  \begin{aligned}
    \mathbf D_t(\omega)
      &=\mathbf D_0(\omega)+t\mathbf E(\omega),
      \qquad 0\le t\le1,\\
      &=\mathbf D_0(\omega)
        \left(\mathbf I+t\mathbf D_0(\omega)^{-1}\mathbf E(\omega)\right).
  \end{aligned}
  \label{eq:response-homotopy}
\end{equation}
If $q_\Gamma<1$, the second factor remains invertible on $\Gamma$ throughout this change. No root can cross the boundary, and the number of enclosed roots, counted with multiplicity, remains unchanged \cite[Lemma~2.1 and Remark~1]{bindelHood2013localization}. Theorem~\ref{thm:app-contour} gives the full argument.

A sufficient test that avoids explicit matrix inversion uses the minimum singular value and the matrix difference:
\begin{equation}
  \mu_\Gamma
  =\inf_{\omega\in\Gamma}
   \sigma_{\min}\!\left(\mathbf D_0(\omega)\right),
  \qquad
  \delta_\Gamma
  =\sup_{\omega\in\Gamma}
   \left\|\mathbf D_1(\omega)-\mathbf D_0(\omega)\right\|_2.
  \label{eq:mu-delta-definitions}
\end{equation}
Here $\delta_\Gamma$ measures the perturbation on the boundary, while $\mu_\Gamma$ measures the distance of $\mathbf D_0$ from singularity there. Since $q_\Gamma\le\delta_\Gamma/\mu_\Gamma$, the condition
\begin{equation}
  \delta_\Gamma<\mu_\Gamma,
  \label{eq:mu-delta-condition}
\end{equation}
is sufficient for $\mathbf D_0$ and $\mathbf D_1$ to have the same number of roots in $U$. The perturbation is then smaller than the boundary margin to singularity. Failure of a sufficient test is inconclusive by itself; it does not identify an extraneous root.

For an individual candidate $\widehat\omega$, take $U=B(\widehat\omega,\rho)$. After independently establishing that $\mathbf D_0$ has exactly one root in this disk, counted with multiplicity, Eq.~\eqref{eq:mu-delta-condition} guarantees a unique simple root $\omega_1$ of $\mathbf D_1$ satisfying
\begin{equation}
  |\omega_1-\widehat\omega|<\rho.
  \label{eq:isolated-root-disk}
\end{equation}
The radius $\rho$ is therefore an error bound relative to the chosen reference matrix. If the disk contains a cluster, the same test preserves its total root count rather than pairing the individual roots (Corollary~\ref{cor:app-localization}).

The extrema in Eq.~\eqref{eq:mu-delta-definitions} are taken over the entire contour. To evaluate the test from discrete samples, partition $\Gamma$ into subarcs $\Gamma_j$, choose a sample point $z_j$ on each subarc, and let $a_j$ bound the arc length from $z_j$ to any point of $\Gamma_j$. Suppose that
\begin{equation}
  \sup_{\omega\in\Gamma_j}\|\mathbf E'(\omega)\|_2\le L_{E,j},
  \qquad
  \sup_{\omega\in\Gamma_j}\|\mathbf D_0'(\omega)\|_2\le L_{0,j},
  \label{eq:contour-derivative-bounds}
\end{equation}
and that the sample values have bounds $u_{E,j}\ge\|\mathbf E(z_j)\|_2$ and $l_{0,j}\le\sigma_{\min}(\mathbf D_0(z_j))$, including numerical evaluation errors. Derivative bounds control how much the matrices can change between a sample point and the rest of its subarc. The resulting continuous bounds are
\begin{equation}
  \begin{aligned}
    \overline\delta_\Gamma
      &=\max_j\left(u_{E,j}+L_{E,j}a_j\right),\\
    \underline\mu_\Gamma
      &=\min_j\left(l_{0,j}-L_{0,j}a_j\right).
  \end{aligned}
  \label{eq:continuous-boundary-enclosures}
\end{equation}
They satisfy $\delta_\Gamma\le\overline\delta_\Gamma$ and $\mu_\Gamma\ge\underline\mu_\Gamma$. Thus the computable condition $\overline\delta_\Gamma<\underline\mu_\Gamma$ establishes Eq.~\eqref{eq:mu-delta-condition} on the continuous contour, including the intervals between samples. Appendix~\ref{app:boundary} supplies the proof, the bounds on numerical arithmetic, and derivative bounds obtained from rational coefficients.

\subsection{Sensitivity analysis}
\label{sec:growth-rate-inference}

The same matrix approximation can produce different frequency errors on different branches. To describe this local sensitivity, let $\widehat\omega$ be a simple root of $\mathbf D_0$ and let $\mathbf v$ and $\mathbf w$ be nonzero right and left null vectors of $\mathbf D_0(\widehat\omega)$. Along the family $\mathbf D_t=\mathbf D_0+t\mathbf E$ in Eq.~\eqref{eq:response-homotopy}, the local root branch $\omega(t)$ with $\omega(0)=\widehat\omega$ satisfies
\begin{equation}
  \left.\frac{\mathrm d\omega}{\mathrm dt}\right|_{t=0}
  =-
  \frac{\mathbf w^*\mathbf E(\widehat\omega)\mathbf v}
       {\mathbf w^*\mathbf D_0'(\widehat\omega)\mathbf v},
  \qquad
  \mathbf w^*\mathbf D_0'(\widehat\omega)\mathbf v\ne0.
  \label{eq:simple-root-sensitivity}
\end{equation}
The numerator selects the part of the matrix perturbation that acts on this mode. The denominator is the frequency derivative projected onto its left and right null vectors. For unit vectors $\mathbf v$ and $\mathbf w$,
\begin{equation}
  |\omega'(0)|
  \le
  \frac{\|\mathbf E(\widehat\omega)\|_2}
       {|\mathbf w^*\mathbf D_0'(\widehat\omega)\mathbf v|}.
  \label{eq:root-sensitivity-bound}
\end{equation}
A small projected derivative permits a larger frequency change for the same matrix error. This explains the branch dependence of root accuracy, but describes only the behavior near $t=0$. The finite error radius at $t=1$ is supplied by the contour test. The derivative formula and its proof are given in Proposition~\ref{prop:app-sensitivity}; its relation to eigenvalue conditioning is discussed in Ref.~\cite[Sec.~2.7, proof of Theorem~2.20]{guttelTisseur2017nep}.

When $\mathbf D_1=\mathbf D_\star$, the boundary difference $\delta_\Gamma$ includes all approximations between the computed and target dispersion matrices. Consider a sequence from $\mathbf D_0$ to $\mathbf D_\star$ in the common matrix representation, changing one approximation at a time. Let $\delta_{\mathrm{fit}}$, $\delta_J$, $\delta_N$, and $\delta_{\mathrm{SVD}}$ be uniform bounds on $\Gamma$ for the changes due to distribution fitting, rational expansion, harmonic truncation, and numerical residue truncation, respectively. Here distribution fitting includes any finite basis representation in velocity space. The triangle inequality gives
\begin{equation}
  \delta_\Gamma
  \le
  \delta_{\mathrm{fit}}+\delta_J+\delta_N+\delta_{\mathrm{SVD}}.
  \label{eq:total-dispersion-error}
\end{equation}
Appendix~\ref{app:error-budget} defines the intermediate matrices and establishes this sum of bounds. Numerical errors in evaluating the contour matrices are included in the sample bounds $u_{E,j}$ and $l_{0,j}$. A comparison with a finite reference $\mathbf D_{h_1}$ controls the difference from that approximation; a bound relative to $\mathbf D_\star$ also requires the remaining truncation error.

Finally, the error radius determines whether the sign of the growth rate is resolved. For the reference root $\omega_1$ enclosed by Eq.~\eqref{eq:isolated-root-disk}, the convention in Eq.~\eqref{eq:plane-wave-convention} gives $\gamma_1=\operatorname{Im}\omega_1$ and
\begin{equation}
  \gamma_1
  \in
  [\operatorname{Im}\widehat\omega-\rho,
   \operatorname{Im}\widehat\omega+\rho].
  \label{eq:growth-rate-interval}
\end{equation}
An interval wholly above zero certifies instability; an interval wholly below zero certifies damping. These conditions are $\operatorname{Im}\widehat\omega-\rho>0$ and $\operatorname{Im}\widehat\omega+\rho<0$, respectively. If the interval contains zero, the chosen radius does not determine the sign. A contour enclosing a cluster gives the same sign conclusion when it lies entirely in one open half-plane (Corollary~\ref{cor:app-growth}).

\section{Numerical examples}
\label{sec:numerical-examples}

\subsection{Experimental setup}
\label{sec:numerical-setup}

We now benchmark BO-T on thirteen test cases from PDRK, BO, and BO-Arbitrary \cite{xieXiao2016pdrk,xie2019bo,xie2025boarbitrary}, covering electrostatic and electromagnetic models, analytical and Hermite--Hermite (HH) fitted distributions, and 1D and 2D wave vector scans. We first test whether the reduced matrix eigenvalue problem reproduces all eigenvalues of the original BO/PDRK matrix and measure the speedup as the cyclotron harmonic number $N$ increases. We then compare the physical dispersion branches selected by BO-T with reference solutions from the exact $Z$ function, published figures, or archived BO data.

For the timing tests, we use eleven magnetized cases; two low-dimensional electrostatic cases (ES1D beam and ES3D loss cone) are omitted from the timing tables because their matrices are already small and matrix reduction gives no speedup. Fixing the plasma parameters, $J$, and the wave vector at the midpoint of each scan, we increase $N$ over a preset grid starting from either the default $N$ or the smallest $N$ with full matrix dimension $d_{\rm BO}\ge 1000$, up to $d_{\rm BO}\le 5500$. At each $N$, we run one warm-up calculation followed by two preliminary paired runs comparing BO and BO-T. Once the speedup reaches 30 and the full and reduced eigenvalue spectra agree (including multiplicities), we perform another warm-up calculation followed by ten paired runs to measure the mean computation time and speedup; if the mean speedup is below 30, we continue to the next $N$. For cases whose speedup stays below 30 up to $d_{\rm BO}\le 5500$, Table~\ref{tab:scale30-source-backend} lists the timing from the two preliminary runs at the largest $N$.

All calculations are performed in Python on a Windows 11 workstation with an AMD Ryzen 7 9800X3D CPU using a single BLAS thread, alternating BO and BO-T in each pair of runs. In each run, BO constructs and solves the full matrix eigenvalue problem, whereas BO-T constructs the residue blocks, reduces them via SVD, solves the reduced eigenvalue problem, and restores the removed pole eigenvalues. For HH cases, BO calls the single-point routine, while BO-T operates on the initialized scan point. Timing includes matrix assembly and the eigenvalue calculation, but excludes file I/O, initialization, and subsequent mode selection. We report means and sample standard deviations of computation times and paired speedup ratios. Across the timing warm-up runs, the full and reduced spectra agree with multiplicity within a normalized tolerance of $10^{-4}$. Among the eleven cases, only the beam mode and parallel electromagnetic modes show a sharp drop in the singular values of $\mathbf Q_g$ at the truncation threshold. In addition to Table~\ref{tab:scale30-source-backend}, Table~\ref{tab:n32-scaling-source-backend} compares the computation times at fixed $N=32$ for three representative cases.

In Figs.~\ref{fig:es1d-beam}--\ref{fig:loss-cone-mirror}, gray dots show all eigenvalues from the BO matrix in the plot window, and colored markers show roots selected by BO-T. After selection, we consider a selected root $\omega_{\mathrm{sel}}$ to match a reference point $\omega_{\mathrm{ref}}$ when $|\omega_{\mathrm{sel}}-\omega_{\mathrm{ref}}|/(1+|\omega_{\mathrm{ref}}|)\leq10^{-3}$, where $\omega_{\mathrm{ref}}$ is taken from the original solver, digitized published curves, or archived data from Refs.~\cite{xieXiao2016pdrk,xie2019bo,xie2025boarbitrary}.

In Table~\ref{tab:scale30-source-backend}, the results for parallel electromagnetic modes, 2D surface, HH firehose, HH ring beam, and HH shell are from two paired runs at the largest $N$, and the remaining six cases are from ten paired runs. The settings column lists $J$, $N$, and the full matrix dimension $d_{\rm BO}$.

\begin{table}[!htbp]
\centering
\footnotesize
\setlength{\tabcolsep}{3pt}
\caption{Computation times for eleven BO/PDRK benchmark cases (mean $\pm$ standard deviation).}
\label{tab:scale30-source-backend}
\begin{tabular}{@{}lcccc@{}}
\toprule
Case & Settings & BO (ms) & BO-T (ms) & Speedup \\
\midrule
Beam mode & $J=8,\;N=8,\;d_{\rm BO}=1641$ & $1008\pm27$ & $30.0\pm1.1$ & $33.62\pm0.59$ \\
Cold plasma & $J=8,\;N=24,\;d_{\rm BO}=2361$ & $3635\pm16$ & $55.3\pm1.3$ & $65.8\pm1.6$ \\
Parallel electromagnetic modes & $J=8,\;N=48,\;d_{\rm BO}=4665$ & $749\pm56$ & $73.3\pm1.5$ & $10.21\pm0.57$ \\
2D surface & $J=8,\;N=48,\;d_{\rm BO}=4665$ & $27612\pm28$ & $2528.3\pm5.1$ & $10.921\pm0.033$ \\
Loss-cone mirror & $J=8,\;N=32,\;d_{\rm BO}=3129$ & $9432\pm66$ & $227\pm15$ & $41.6\pm2.5$ \\
Perpendicular beam & $J=8,\;N=60,\;d_{\rm BO}=2961$ & $9591\pm84$ & $76.4\pm2.1$ & $125.6\pm3.6$ \\
Analytical ring beam & $J=8,\;N=48,\;d_{\rm BO}=4665$ & $39430\pm660$ & $493.1\pm8.9$ & $80.0\pm1.6$ \\
HH firehose & $J=24,\;N=16,\;d_{\rm BO}=4764$ & $31660\pm490$ & $3127\pm51$ & $10.13\pm0.32$ \\
HH ICE & $J=4,\;N=64,\;d_{\rm BO}=4659$ & $25760\pm350$ & $569.4\pm7.5$ & $45.25\pm0.90$ \\
HH ring beam & $J=12,\;N=32,\;d_{\rm BO}=4692$ & $38720\pm36$ & $1433.8\pm7.6$ & $27.01\pm0.17$ \\
HH shell & $J=24,\;N=12,\;d_{\rm BO}=5415$ & $45500\pm1200$ & $4720\pm220$ & $9.64\pm0.20$ \\
\bottomrule
\end{tabular}
\end{table}
\FloatBarrier

\subsection{Computational performance}
\label{sec:source-backend-results}

As shown in Table~\ref{tab:scale30-source-backend}, six of the eleven cases reach a mean speedup above 30: the beam mode ($33.62\pm0.59$), cold plasma ($65.8\pm1.6$), loss-cone mirror ($41.6\pm2.5$), perpendicular beam ($125.6\pm3.6$), analytical ring beam ($80.0\pm1.6$), and HH ICE ($45.25\pm0.90$). The remaining five cases have speedups between $9.64\pm0.20$ and $27.01\pm0.17$ at the largest $N$ tested ($d_{\rm BO}\le 5500$). Table~\ref{tab:n32-scaling-source-backend} further compares three cases at fixed $N=32$, where the speedup is $144.0\pm4.2$ for the cold plasma, $28.20\pm0.44$ for the analytical ring beam ($d_{\rm BO}=3129$), and $24.4\pm2.9$ for the beam mode ($d_{\rm BO}=6249$). This difference reflects the smaller rank $r_g$ of the residue blocks in the cold-plasma and ring-beam cases. Note that evaluating the contour integral check at twelve representative scan points is currently slower than solving the full eigenvalue problem, so the speedups in Tables~\ref{tab:scale30-source-backend} and \ref{tab:n32-scaling-source-backend} measure only the matrix reduction of Section~\ref{sec:response-preserving-reduction}.

\begin{table}[!htbp]
\centering
\footnotesize
\setlength{\tabcolsep}{3pt}
\caption{Computation times at fixed $N=32$ over ten paired runs (mean $\pm$ standard deviation).}
\label{tab:n32-scaling-source-backend}
\begin{tabular}{@{}lcccc@{}}
\toprule
Case & Settings & BO (ms) & BO-T (ms) & Speedup \\
\midrule
Beam mode & $J=8,\ N=32,\ d_{\rm BO}=6249$ & $2481\pm 49$ & $103\pm 14$ & $24.4\pm 2.9$ \\
Cold plasma & $J=8,\ N=32,\ d_{\rm BO}=3129$ & $9544\pm 32$ & $66.3\pm 2.1$ & $144.0\pm 4.2$ \\
Analytical ring beam & $J=8,\ N=32,\ d_{\rm BO}=3129$ & $12740\pm 150$ & $451.7\pm 3.6$ & $28.20\pm 0.44$ \\
\bottomrule
\end{tabular}
\end{table}
\FloatBarrier

\subsection{Dispersion examples}
\label{sec:dispersion-examples}

\begin{figure}[!htbp]
  \centering
  \includegraphics[width=\linewidth]{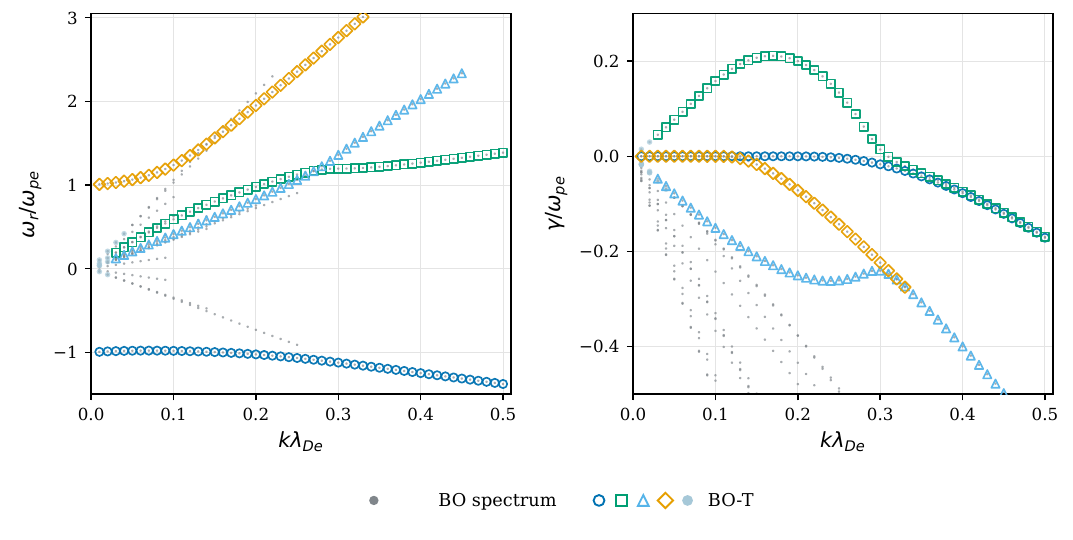}
  \caption{Electron bump-on-tail instability \cite{xieXiao2016pdrk}: BO-T solutions (colored markers) and all eigenvalues of the BO matrix (gray dots), reproducing the three branches with the largest imaginary parts of the matrix method.}
  \label{fig:es1d-beam}
\end{figure}

\paragraph{Electrostatic bump-on-tail instability.}
Figure~\ref{fig:es1d-beam} shows the electron bump-on-tail instability from Ref.~\cite{xieXiao2016pdrk}, calculated here using the ES3D solver with parallel propagation ($k=k_\parallel$). Across all 40 wave numbers in Fig.~2 of Ref.~\cite{xieXiao2016pdrk}, BO-T reproduces the three branches with the largest imaginary parts of the $J$-pole matrix system. The unstable branch agrees with the exact $Z(\zeta)$ contour solution across all 50 scan points (including all 30 unstable points), while the two damped branches correspond to solutions of the $J$-pole approximation in PDRK \cite{xieXiao2016pdrk}.

\begin{figure}[!htbp]
  \centering
  \includegraphics[width=\linewidth]{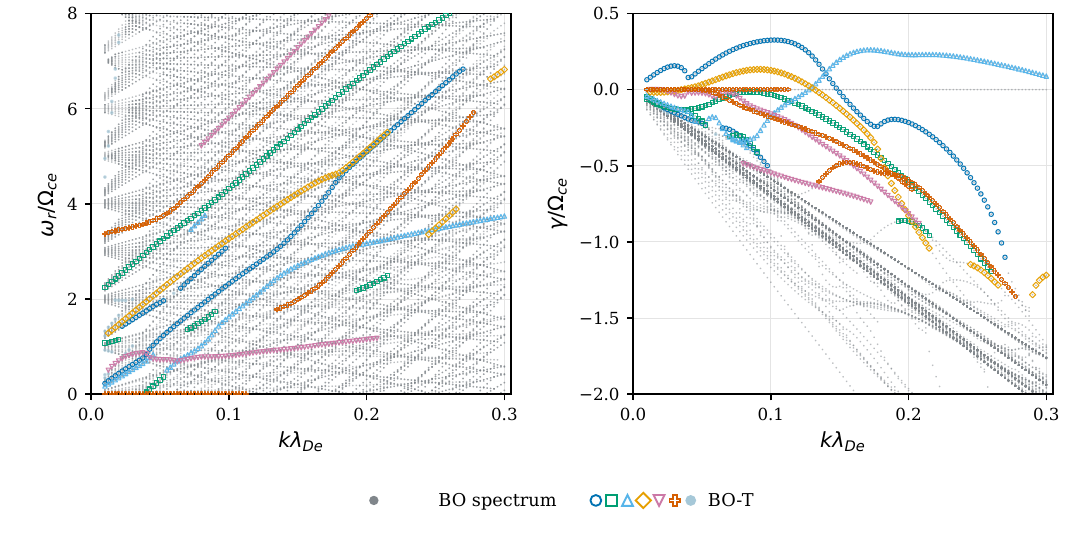}
  \caption{Electromagnetic ring-beam instability with an analytical distribution \cite{xie2019bo,umedaEtAl2012ringbeam}: BO-T recovers the unstable ranges of all three reference BO branches.}
  \label{fig:umeda-ring-beam}
\end{figure}

\paragraph{Analytical ring-beam instability.}
Figure~\ref{fig:umeda-ring-beam} shows the electromagnetic ring-beam instability of Ref.~\cite{umedaEtAl2012ringbeam} calculated with the analytical ring-beam model in BO \cite{xie2019bo}. Compared with the three reference branches from Ref.~\cite{xie2019bo}, BO-T recovers all $163/163$ unstable points and $292/351$ points overall; the 59 missing points lie on the heavily damped parts of the branches. Within the plotted frequency and growth rate ranges, BO-T also retains 512 points on other weakly damped or oscillatory branches of the BO matrix.

\begin{figure}[!htbp]
  \centering
  \includegraphics[width=\linewidth]{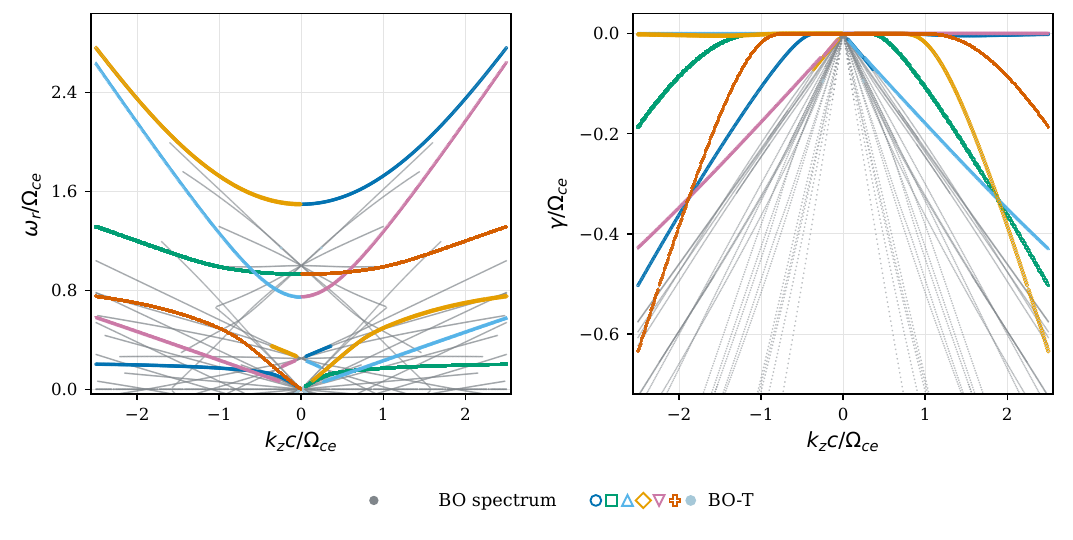}
  \caption{Parallel electromagnetic modes \cite{xie2019bo}: BO-T solutions match all five published dispersion branches for both signs of $k_z c/\Omega_{ce}$ ($2498/2500$ points).}
  \label{fig:signed-em}
\end{figure}

\paragraph{Parallel electromagnetic modes.}
Figure~\ref{fig:signed-em} shows parallel-propagating electromagnetic waves for both positive and negative $k_z c/\Omega_{ce}$ \cite{xie2019bo} (the timing test in Table~\ref{tab:scale30-source-backend} uses the midpoint of the $k_z>0$ scan with $N=48$). Comparing the selected solutions with five published dispersion curves (digitized and interpolated to 500 wave numbers, giving 2500 reference points) shows agreement at $2498/2500$ points within $10^{-3}$. The only two missing points are at $k_z c/\Omega_{ce}=\pm2.25$, where the eigenvalues are present in the spectrum but removed by the mode selector. The selected set also contains 548 points belonging to other modes in the plotted window.

\begin{figure}[!htbp]
  \centering
  \includegraphics[width=\linewidth]{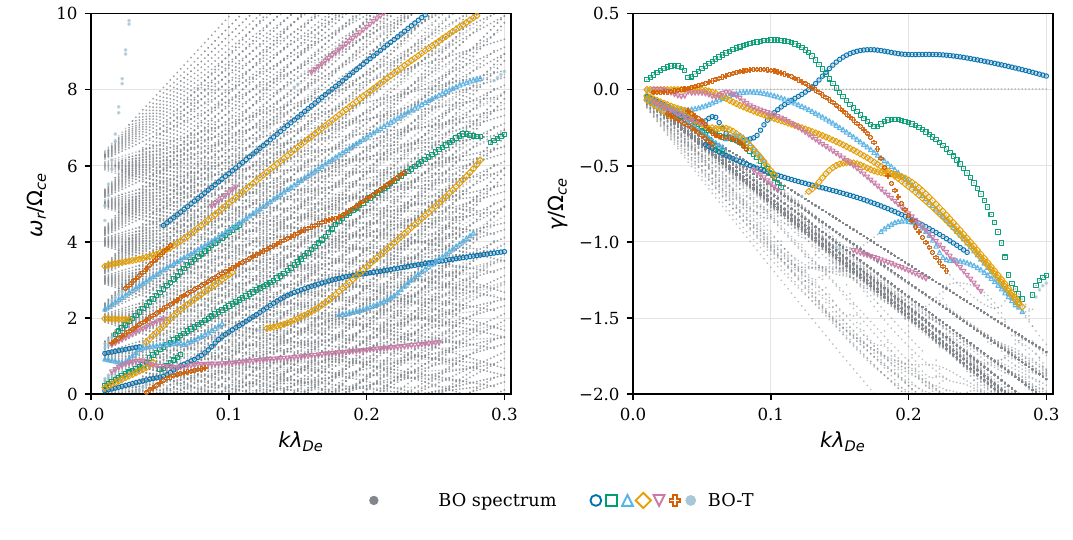}
  \caption{Ring-beam instability using an HH expansion \cite{xie2025boarbitrary,umedaEtAl2012ringbeam}: comparing the $(J,N)=(12,6)$ spectrum against $J=24$ retains all $163/163$ unstable points and $324/351$ reference points overall.}
  \label{fig:hh-ring-beam}
\end{figure}

\paragraph{Ring-beam instability with an HH expansion.}
Figure~\ref{fig:hh-ring-beam} shows the same ring-beam instability solved using the Hermite--Hermite (HH) expansion in BO-Arbitrary \cite{xie2025boarbitrary,umedaEtAl2012ringbeam} across 117 wave numbers with $(J,N)=(12,6)$ (whereas Table~\ref{tab:scale30-source-backend} times $(J,N)=(12,32)$). Using $J=8$ in the comparison test recovers $120/163$ unstable points and $236/351$ points overall; increasing the comparison order to $J=24$ recovers all $163/163$ unstable points and $324/351$ reference points in total, with the remaining 27 points on damped segments.

\begin{figure}[!htbp]
  \centering
  \includegraphics[width=\linewidth]{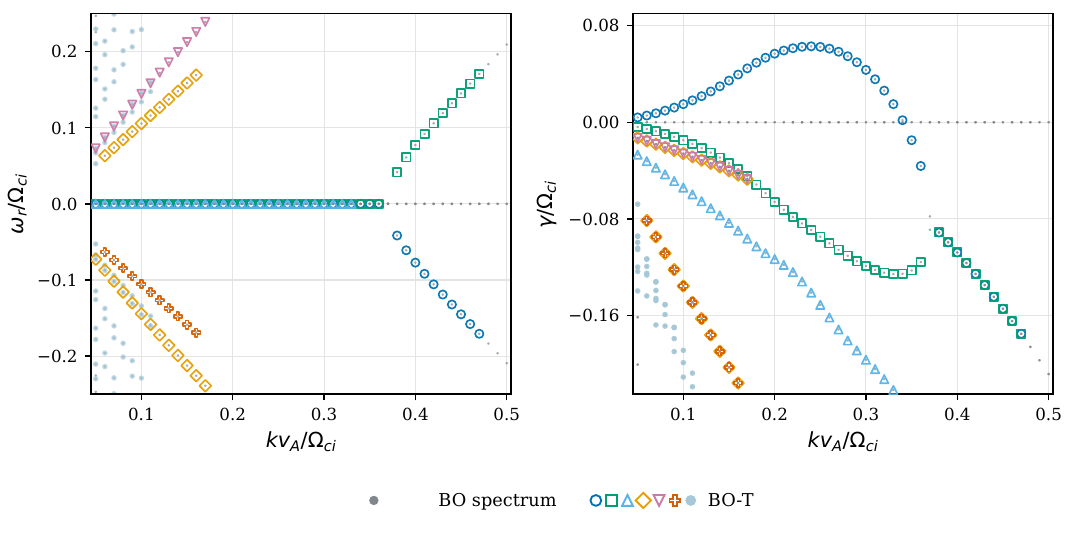}
  \caption{Oblique firehose instability \cite{astfalkJenko2017leopard,xie2025boarbitrary}: BO-T recovers all $30/30$ unstable points and $42/46$ reference points.}
  \label{fig:firehose}
\end{figure}

\Needspace{11\baselineskip}
\paragraph{Oblique firehose instability.}
Figure~\ref{fig:firehose} shows the oblique firehose instability across 46 wave numbers with $N=2$ (compared with $N=16$ in Table~\ref{tab:scale30-source-backend}), using the BO-Arbitrary solution as reference \cite{astfalkJenko2017leopard,xie2025boarbitrary}. The dimensionless wave number plotted here is $kv_A/\Omega_{ci}$ (labeled as $k\lambda_{Di}$ in Ref.~\cite{xie2025boarbitrary}). BO-T recovers all $30/30$ unstable points and $42/46$ reference points overall, missing four damped points. At 38 of the 46 wave numbers, all eigenvalues of the full and reduced matrices match directly within $10^{-4}$, while at the other eight wave numbers, eleven ill-conditioned multiple-eigenvalue clusters are verified by their matrix residuals. At a single scan point, the fast and exhaustive local selection checks differ on two candidate roots.

\begin{figure}[!htbp]
  \centering
  \includegraphics[width=\linewidth]{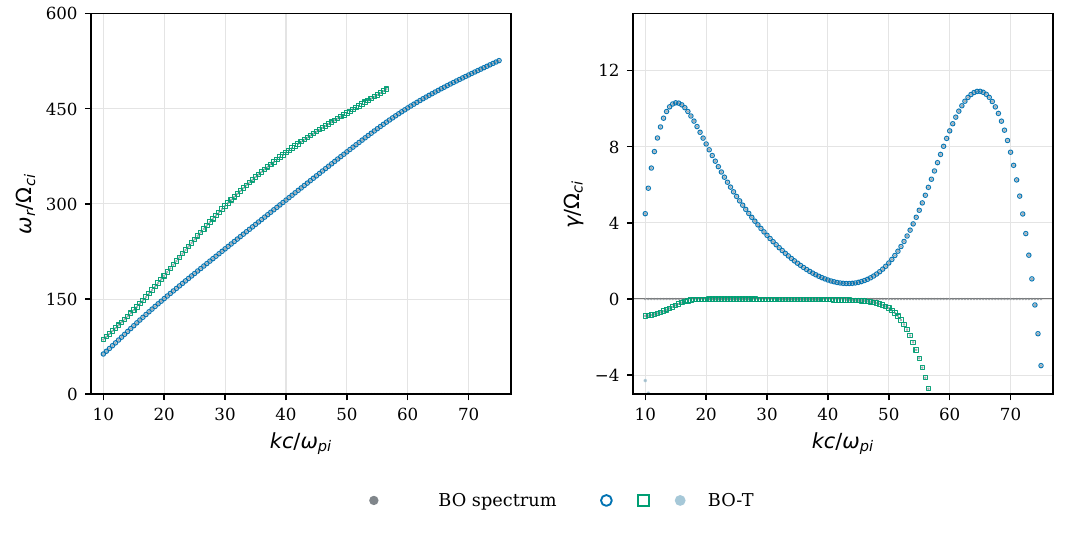}
  \caption{Perpendicular beam instability \cite{muschiettiLembege2017shock,xie2019bo}: BO-T reproduces both growth rate peaks and matches $123/131$ reference points.}
  \label{fig:muschietti-beam}
\end{figure}

\Needspace{8\baselineskip}
\paragraph{Perpendicular beam instability.}
Figure~\ref{fig:muschietti-beam} shows the perpendicular beam instability with high-order cyclotron harmonics \cite{muschiettiLembege2017shock,xie2019bo}. BO-T reproduces both peaks of the growth rate and the real frequency curve, matching $123/131$ reference points from Ref.~\cite{xie2019bo} within $10^{-3}$. The eight differing points at small $k$ are already present between the recomputed full BO spectrum and the archived file prior to mode selection.

\begin{figure}[!htbp]
  \centering
  \includegraphics[width=\linewidth]{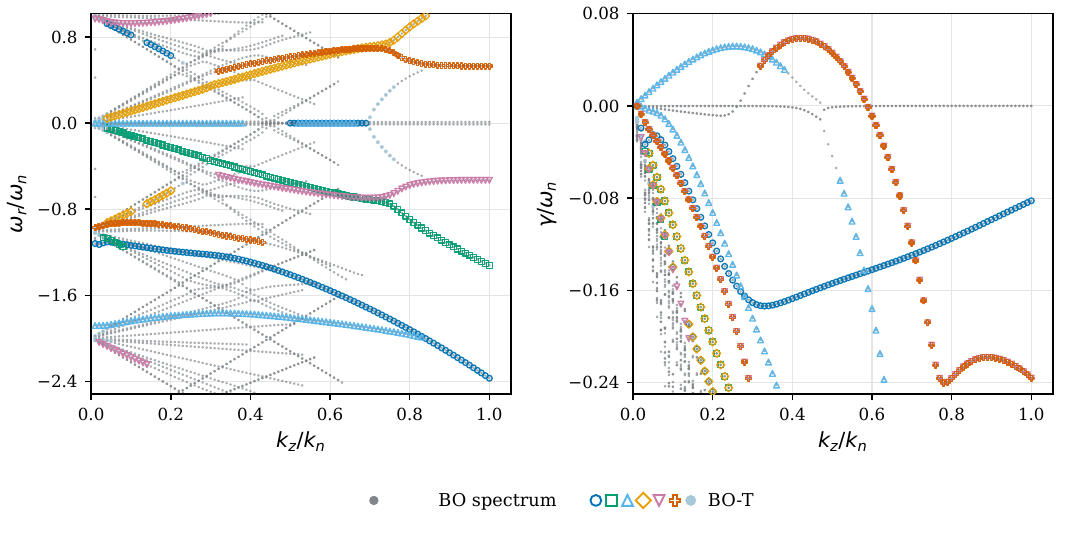}
  \caption{Loss-cone mirror instability \cite{xie2019bo}: BO-T captures both growth peaks and $298/400$ reference points.}
  \label{fig:loss-cone-mirror}
\end{figure}

\paragraph{Loss-cone mirror instability.}
Figure~\ref{fig:loss-cone-mirror} shows the loss-cone mirror instability from Ref.~\cite{xie2019bo}. BO-T reproduces both growth peaks and matches $298/400$ points across the four reference branches in Ref.~\cite{xie2019bo}, with two gaps of five points in the unstable range. Checking the full BO spectrum shows that 100 of the 102 unmatched reference points do not have a corresponding eigenvalue in the recomputed full BO matrix within $10^{-3}$, and only two points are removed by the selector.

\subsection{Other test cases and discussion}
\label{sec:numerical-scope}

The remaining six test cases are used mainly to test the timing and accuracy of the matrix reduction. Specifically, the cold-plasma, 2D Bernstein dispersion surface, and HH ion cyclotron emission (ICE) cases use different parameters or wave-number grids from the published figures, and the ES3D loss-cone case is a small-matrix test without an external reference curve. In the whistler beam-mode case, the current selector retains $232/400$ archived points, while in the HH shell case it misses part of the unstable peak and shows differences between the fast and exhaustive selection checks. Overall, the numerical tests show that the SVD-based matrix reduction reliably accelerates the BO/PDRK eigenvalue calculation for large $N$ without changing the spectrum, while the contour-based criterion provides root bounds relative to the chosen reference dispersion relation.
\FloatBarrier

\section{Summary and discussion}
\label{sec:conclusion}

In this paper, we have developed BO-T to accelerate the matrix eigenvalue calculation and separate physical modes from artificial solutions in the kinetic dispersion relation solver BO. By grouping identical poles and applying SVD to factorize their combined residue matrices, BO-T reduces the dimension of the linearized eigenvalue problem without changing the rational dispersion relation. Timing benchmarks on eleven magnetized plasma cases show speedups of about 10 to 140 over the unreduced solver, with six cases exceeding a speedup of 30. In addition, the mode selector recovers the main unstable branches and growth-rate peaks in the bump-on-tail, ring-beam, firehose, perpendicular beam, and mirror benchmarks, as well as the parallel electromagnetic dispersion curves.

We note that the speedups reported in Tables~\ref{tab:scale30-source-backend} and \ref{tab:n32-scaling-source-backend} include only matrix assembly and eigenvalue calculation. Evaluating the contour-based root localization directly is still computationally expensive, and in several heavily damped regimes only part of the reference branch is retained. The contour criterion provides root bounds when bounds over the entire contour are available. Further improving the efficiency of the contour evaluation and mode selection for large parameter scans will be left for future work.

\section*{Acknowledgement}

This research is supported by Grant JIH2325037Y from Fudan University. The computations in this research were performed using the CFFF platform of Fudan University.

\section*{Code availability}

The BO-T source code and reproduction materials are available at
\url{https://github.com/FusionAlpha/botian}.

\FloatBarrier


\clearpage
\appendix
\setlength{\parskip}{0pt}
\setlength{\abovedisplayskip}{8pt plus 2pt minus 2pt}
\setlength{\belowdisplayskip}{8pt plus 2pt minus 2pt}
\AddToHook{cmd/subsection/before}{\Needspace{8\baselineskip}}
\numberwithin{equation}{section}
\section{Proofs of matrix reduction and root localization}
\label{app:proofs}

This appendix proves the reduction and localization statements in Section~\ref{sec:methodology}. The physical parameters and wave vector are fixed throughout. All matrices are over $\mathbb C$, $\|\cdot\|_2$ denotes the spectral norm, and $\mathbf X^*$ denotes conjugate transpose. Algebraic multiplicity means the order of a determinant zero. Identities hold in exact arithmetic, whereas numerical bounds must enclose the corresponding exact quantities.

\subsection{Exact reduction and spectral multiplicities}
\label{app:exact-reduction}

Assume the original linear matrix function has the form in Eq.~\eqref{eq:full-grouped-pencil}, with distinct pole locations $\xi_g$ of the auxiliary block, positive integers $b_g$, constant blocks $\mathbf B,\mathbf C,\mathbf F,\mathbf M$, and $\det\mathcal L_h^{\mathrm{full}}\not\equiv0$. Define $\mathbf Q_g=\mathbf C_g\mathbf B_g$ and $r_g=\operatorname{rank}\mathbf Q_g$. Choose exact rank factorizations $\mathbf Q_g=\mathbf L_g\mathbf R_g$ and construct $\mathcal L_h^{\mathrm{redu}}$ as in Eqs.~\eqref{eq:reduced-factors}--\eqref{eq:reduced-grouped-pencil}. A group with $r_g=0$ contributes no reduced auxiliary rows or columns. Let $\mathcal P=\{\xi_1,\ldots,\xi_G\}$ and $r=\sum_g r_g$.

\begin{theorem}[Preservation of the Schur complement and determinant identity]
\label{thm:exact-reduction}
Under these assumptions, the original and reduced eigenvalue problems have the same Schur complement
\begin{equation}
  \mathbf S_h(\omega)=\mathbf F-\omega\mathbf M
       +\sum_{g=1}^G\frac{\mathbf Q_g}{\omega-\xi_g},
  \qquad \omega\notin\mathcal P.
  \label{eq:app-schur}
\end{equation}
The reduced linear matrix function is regular, and the polynomial identity
\begin{equation}
  \det\mathcal L_h^{\mathrm{full}}(\omega)
  =\det\mathcal L_h^{\mathrm{redu}}(\omega)
       \prod_{g=1}^G(\xi_g-\omega)^{b_g-r_g}
  \label{eq:app-determinant}
\end{equation}
holds for all $\omega\in\mathbb C$. Consequently, eigenvalues away from $\mathcal P$ retain their algebraic multiplicities, and at each $\xi_g$,
\begin{equation}
  \operatorname{mult}_{\xi_g}\mathcal L_h^{\mathrm{full}}
  =\operatorname{mult}_{\xi_g}\mathcal L_h^{\mathrm{redu}}+b_g-r_g.
  \label{eq:app-pole-multiplicity}
\end{equation}
The total multiplicity at infinity is also unchanged when both linear matrix functions are regarded as matrix polynomials of grade one.
\end{theorem}

\begin{proof}
For $\omega\notin\mathcal P$, the auxiliary block is invertible. Its elimination gives
\[
  \mathbf F-\omega\mathbf M
       -\mathbf C(\boldsymbol\Xi-\omega\mathbf I_b)^{-1}\mathbf B
  =\mathbf F-\omega\mathbf M
       +\sum_g\frac{\mathbf C_g\mathbf B_g}{\omega-\xi_g}.
\]
Replacing $\mathbf C_g\mathbf B_g$ by $\mathbf L_g\mathbf R_g$ gives exactly the reduced Schur complement. The block determinant formula therefore yields
\[
  \det\mathcal L_h^{\mathrm{full}}
    =\prod_g(\xi_g-\omega)^{b_g}\det\mathbf S_h,
  \qquad
  \det\mathcal L_h^{\mathrm{redu}}
    =\prod_g(\xi_g-\omega)^{r_g}\det\mathbf S_h
\]
away from $\mathcal P$. Since $r_g\le b_g$, both sides of Eq.~\eqref{eq:app-determinant} are polynomials. They agree outside a finite set and hence are identical. Regularity of the original linear matrix function then implies regularity of the reduced one. Taking orders of zeros proves the finite multiplicity statements; distinctness of the $\xi_g$ ensures that the other factors are nonzero at $\xi_g$.

For a regular $n\times n$ linear matrix polynomial $\mathcal L(\omega)=\mathbf A-\omega\mathbf B$, its reversal with grade one is $\operatorname{rev}_1\mathcal L(z)=z\mathcal L(1/z)=z\mathbf A-\mathbf B$. If $d=\deg\det\mathcal L$, then $\det[z\mathcal L(1/z)]=z^n\det\mathcal L(1/z)$ has a zero of order $n-d$ at $z=0$. This is the total multiplicity at infinity. Equation~\eqref{eq:app-determinant} increases the determinant degree by $b-r$, exactly the increase in matrix dimension, so $n-d$ is the same for both eigenvalue problems.
\end{proof}

Away from $\mathcal P$, the eigenvectors are also preserved. If $\mathbf S_h(\lambda)\mathbf v=0$ and $\mathbf v\ne0$, the full and reduced eigenvectors are respectively
\begin{equation}
  \begin{bmatrix}
    (\lambda\mathbf I_b-\boldsymbol\Xi)^{-1}\mathbf B\mathbf v\\ \mathbf v
  \end{bmatrix},
  \qquad
  \begin{bmatrix}
    (\lambda\mathbf I_r-\boldsymbol\Xi_r)^{-1}\mathbf R\mathbf v\\ \mathbf v
  \end{bmatrix}.
  \label{eq:app-eigenvector-lifting}
\end{equation}
Substitution proves both identities. Conversely, the first block row determines the auxiliary components uniquely, and a nonzero eigenvector cannot have a zero retained component. Thus these maps are bijections between the corresponding kernels for $\lambda\notin\mathcal P$.

\begin{remark}[Scope at poles and at singular normalizations]
\label{rem:app-poles}
Equation~\eqref{eq:app-pole-multiplicity} counts algebraic multiplicity, not Jordan blocks. Semisimplicity of $\boldsymbol\Xi$ does not imply semisimplicity of an eigenvalue of the coupled generalized eigenvalue problem at $\xi_g$. Nor does an eigenvalue at $\xi_g$ necessarily correspond to a root of the dispersion relation there. Theorem~\ref{thm:exact-reduction} applies to simple poles and does not give a rank formula for higher-order poles.
\end{remark}

To transfer multiplicities from $\mathbf S_h$ to a smaller dispersion matrix $\mathbf D_h$, the subsequent eliminations must also be regular on the domain of interest. Specifically, suppose on an open set $V\subset\mathbb C\setminus\mathcal P$ there are analytic, invertible matrices $\mathbf U,\mathbf V$ such that
\begin{equation}
  \mathbf U(\omega)\mathbf S_h(\omega)\mathbf V(\omega)
       =\operatorname{diag}(\mathbf K(\omega),\mathbf D_h(\omega)),
  \qquad \det\mathbf K(\omega)\ne0\quad(\omega\in V).
  \label{eq:app-regular-elimination}
\end{equation}
Here the empty block $\mathbf K$ is permitted, with determinant one. Taking determinants shows that $\det\mathbf S_h$ and $\det\mathbf D_h$ differ by an analytic, nonvanishing factor; their zeros therefore have identical orders on $V$. This covers invertible block eliminations and nonsingular matrix scalings. Multiplication by a factor that vanishes at zero frequency, for example, does not meet this condition at that frequency.

\begin{proposition}[Minimal auxiliary dimension for the strictly proper rational matrix]
\label{prop:app-minimality}
The strictly proper rational matrix
\[
  \mathbf H(\omega)=\sum_g\frac{\mathbf Q_g}{\omega-\xi_g}
\]
has a standard state-space realization of dimension $r=\sum_g r_g$, and no realization of the form $\mathbf C_a(\omega\mathbf I_n-\mathbf A_a)^{-1}\mathbf B_a$ has smaller dimension.
\end{proposition}

\begin{proof}
The factors already constructed give $\mathbf H=\mathbf L(\omega\mathbf I_r-\boldsymbol\Xi_r)^{-1}\mathbf R$. For any other realization, decompose the state space into generalized eigenspaces of $\mathbf A_a$. The resolvent on the generalized eigenspace at $\xi_g$ has Laurent expansion
\[
  \frac{\mathbf P_g}{\omega-\xi_g}
  +\frac{(\mathbf A_a-\xi_g\mathbf I)\mathbf P_g}{(\omega-\xi_g)^2}
  +\cdots,
\]
where $\mathbf P_g$ is the spectral projector; set $\mathbf P_g=0$ if $\xi_g$ is not an eigenvalue. Equality of the rational matrix functions implies equality of Laurent coefficients, so $\mathbf Q_g=\mathbf C_a\mathbf P_g\mathbf B_a$ and $r_g\le\operatorname{rank}\mathbf P_g$. The generalized eigenspaces for distinct values of $\xi_g$ form a direct sum. Hence $\sum_g r_g\le\sum_g\operatorname{rank}\mathbf P_g\le n$. The constructed realization attains this lower bound.
\end{proof}

This minimality concerns only $\mathbf H$. It does not assert that $m+r$ is the smallest possible dimension after further reductions of the retained field equations.

\subsection{Error introduced by numerical residue truncation}
\label{app:truncation}

\begin{proposition}[Error bound for approximate factors]
\label{prop:app-truncation}
Keep the pole locations $\xi_g$ and the block $\mathbf F-\omega\mathbf M$ fixed. Let $\widetilde{\mathbf Q}_g$ be approximate residues, and define $\widetilde{\mathbf S}_h$ by replacing $\mathbf Q_g$ with $\widetilde{\mathbf Q}_g$ in Eq.~\eqref{eq:app-schur}. If $\tau_g\ge\|\widetilde{\mathbf Q}_g-\mathbf Q_g\|_2$, then at every $\omega\notin\mathcal P$,
\begin{equation}
  \|\widetilde{\mathbf S}_h(\omega)-\mathbf S_h(\omega)\|_2
     \le\sum_g\frac{\tau_g}{|\omega-\xi_g|}.
  \label{eq:app-truncation-pointwise}
\end{equation}
On any compact contour $\Gamma$ disjoint from $\mathcal P$, writing $d_g=\min_{\omega\in\Gamma}|\omega-\xi_g|>0$ gives
\begin{equation}
  \sup_{\omega\in\Gamma}\|\widetilde{\mathbf S}_h-\mathbf S_h\|_2
     \le\sum_g\frac{\tau_g}{d_g}.
  \label{eq:app-truncation-uniform}
\end{equation}
\end{proposition}

\begin{proof}
Subtracting the two partial-fraction expansions gives $\widetilde{\mathbf S}_h-\mathbf S_h=\sum_g(\widetilde{\mathbf Q}_g-\mathbf Q_g)/(\omega-\xi_g)$. Apply the triangle inequality and then the lower bounds $d_g$.
\end{proof}

For an exact SVD of $\mathbf Q_g$, truncation after $k_g\in\{0,\ldots,m\}$ singular values has defect $\sigma_{k_g+1}(\mathbf Q_g)$, since unitary multiplication preserves the spectral norm of the discarded diagonal block; set $\sigma_{m+1}(\mathbf Q_g)=0$ when all singular values are retained. If the stored factors additionally differ from that exact truncated matrix by at most $\eta_g$, a valid choice is $\tau_g=\sigma_{k_g+1}(\mathbf Q_g)+\eta_g$. This bound must also cover any error in forming the original residue. A numerical rank threshold alone does not establish exact rank or provide $\tau_g$. Even when no nonzero singular value is discarded, evaluation near a pole can amplify small factorization errors through $d_g^{-1}$.

The bound above is for the Schur complement. If the comparison is made after another elimination or scaling, the error must be propagated to that same representation. For fixed left and right scalings $\mathbf U,\mathbf V$, for example, the difference is bounded by $\|\mathbf U\|_2\|\widetilde{\mathbf S}_h-\mathbf S_h\|_2\|\mathbf V\|_2$. Eliminations involving perturbed inverse blocks require additional inverse bounds. Equation~\eqref{eq:app-determinant} holds for exact factorization and does not automatically hold for truncated nonzero singular values.

\subsection{Contour comparison and local root counts}
\label{app:contour}

Let $U$ be a bounded Jordan domain with piecewise smooth, positively oriented boundary $\Gamma$. Let $\mathbf D_0,\mathbf D_1$ be $p\times p$ matrix functions analytic on an open neighborhood of $\overline U$, so that neither matrix has poles in $\overline U$, and assume $\mathbf D_0$ is nonsingular on $\Gamma$. Here $\mathbf D_0$ is the dispersion matrix used to generate the candidate roots and $\mathbf D_1$ is the chosen reference matrix, with the choices specified in Section~\ref{sec:root-localization}. Define $\mathbf E=\mathbf D_1-\mathbf D_0$, and let $N(\mathbf D;U)$ denote the number of eigenvalues of $\mathbf D$ in $U$, counting multiplicity.

\begin{theorem}[Preservation of the local root count]
\label{thm:app-contour}
If
\begin{equation}
  q_\Gamma=\sup_{\omega\in\Gamma}
    \|\mathbf D_0(\omega)^{-1}\mathbf E(\omega)\|_2<1,
  \label{eq:app-relative-test}
\end{equation}
then $\mathbf D_1$ is nonsingular on $\Gamma$ and
\begin{equation}
  N(\mathbf D_1;U)=N(\mathbf D_0;U).
  \label{eq:app-equal-count}
\end{equation}
In particular, $\delta_\Gamma<\mu_\Gamma$, with $\delta_\Gamma$ and $\mu_\Gamma$ defined in Eq.~\eqref{eq:mu-delta-definitions}, is sufficient.
\end{theorem}

\begin{proof}
For $0\le t\le1$, define $\mathbf D_t=\mathbf D_0+t\mathbf E$. On $\Gamma$, $\mathbf D_t=\mathbf D_0(\mathbf I+t\mathbf D_0^{-1}\mathbf E)$, and the second factor is invertible by its convergent Neumann series. Thus $f_t=\det\mathbf D_t$ never vanishes on $\Gamma$ and is not identically zero. Since $\mathbf D_t$ is analytic on a neighborhood of $\overline U$ and nonsingular on $\Gamma$, it has finitely many eigenvalues in $U$. The argument principle and Jacobi's determinant formula give
\begin{equation}
  N(\mathbf D_t;U)
    =\frac{1}{2\pi\mathrm i}\int_\Gamma\frac{f_t'(\omega)}{f_t(\omega)}\,\mathrm d\omega
    =\frac{1}{2\pi\mathrm i}\int_\Gamma
       \operatorname{tr}(\mathbf D_t^{-1}\mathbf D_t')\,\mathrm d\omega.
  \label{eq:app-argument-principle}
\end{equation}
The integrand is continuous in $t$ on the compact set $[0,1]\times\Gamma$. The integral is therefore a continuous, integer-valued function of $t$, hence constant. This proves Eq.~\eqref{eq:app-equal-count}. Finally, compactness and boundary nonsingularity give $\mu_\Gamma>0$, and
\[
  q_\Gamma\le
  \sup_{\omega\in\Gamma}
     \frac{\|\mathbf E(\omega)\|_2}{\sigma_{\min}(\mathbf D_0(\omega))}
  \le\frac{\delta_\Gamma}{\mu_\Gamma}.\qedhere
\]
\end{proof}

The homotopy argument is the matrix localization principle of Bindel and Hood~\cite[Lemma~2.1 and Remark~1]{bindelHood2013localization}, applied here to two specified dispersion matrices. Failure of the strict inequality is inconclusive: it does not imply different root counts.

\begin{corollary}[Isolated roots, clusters, and empty regions]
\label{cor:app-localization}
Under Theorem~\ref{thm:app-contour}, if $U=B(\widehat\omega,\rho)$ and the independently established root count is $N(\mathbf D_0;U)=1$, then $\mathbf D_1$ has exactly one root $\omega_1$ in $U$. It is simple and satisfies $|\omega_1-\widehat\omega|<\rho$. More generally, a root count $q\ge0$ for the computed dispersion matrix gives exactly $q$ roots of the reference dispersion matrix counted with multiplicity; $q=0$ certifies an empty region.
\end{corollary}

\begin{proof}
The counts agree by Theorem~\ref{thm:app-contour}. Each zero of a nonzero holomorphic determinant has positive integer multiplicity, so a total count of one entails one zero of order one. The distance bound is the definition of the disk. The other cases follow from the same count identity.
\end{proof}

Here $\widehat\omega$ need not be an exact root of $\mathbf D_0$; what is required is that $N(\mathbf D_0;U)=1$, which must be verified alongside the boundary inequality. For $q>1$, Theorem~\ref{thm:app-contour} guarantees only that the total multiplicity inside $U$ is preserved.

\subsection{From boundary samples to continuous enclosures}
\label{app:boundary}

Partition $\Gamma$ into finitely many piecewise smooth arcs $\Gamma_j$, with a sample point $z_j\in\Gamma_j$. Let $\ell_j(z_j,\omega)$ be the length of the subarc within $\Gamma_j$ joining $z_j$ to $\omega$, and choose
\begin{equation}
  a_j\ge\sup_{\omega\in\Gamma_j}\ell_j(z_j,\omega).
  \label{eq:app-arc-radius}
\end{equation}
For a circular arc of radius $\rho$ with angular half-width $\alpha_j$ and a midpoint sample, one can take $a_j=\rho\alpha_j$. Assume that the derivatives on each subarc satisfy
\begin{equation}
  \sup_{\omega\in\Gamma_j}\|\mathbf E'(\omega)\|_2\le L_{E,j},
  \qquad
  \sup_{\omega\in\Gamma_j}\|\mathbf D_0'(\omega)\|_2\le L_{0,j}.
  \label{eq:boundary-derivative-bounds}
\end{equation}
At each sample point $z_j$, suppose the bounds
\begin{equation}
  u_{E,j}\ge\|\mathbf E(z_j)\|_2,
  \qquad
  l_{0,j}\le\sigma_{\min}(\mathbf D_0(z_j)).
  \label{eq:app-certified-nodes}
\end{equation}
hold.

\begin{proposition}[Contour bound from sample points]
\label{prop:app-boundary}
Define
\begin{equation}
  \overline\delta_\Gamma=\max_j(u_{E,j}+L_{E,j}a_j),
  \qquad
  \underline\mu_\Gamma=\min_j(l_{0,j}-L_{0,j}a_j).
  \label{eq:app-continuous-bounds}
\end{equation}
Then $\delta_\Gamma\le\overline\delta_\Gamma$ and $\mu_\Gamma\ge\underline\mu_\Gamma$. In particular, $\overline\delta_\Gamma<\underline\mu_\Gamma$ certifies the hypothesis of Theorem~\ref{thm:app-contour}.
\end{proposition}

\begin{proof}
For a matrix function $\mathbf A$ and $\omega\in\Gamma_j$, integration along the subarc gives
\[
  \mathbf A(\omega)-\mathbf A(z_j)
     =\int_{z_j}^{\omega}\mathbf A'(z)\,\mathrm dz,
  \qquad
  \|\mathbf A(\omega)-\mathbf A(z_j)\|_2
     \le a_j\sup_{z\in\Gamma_j}\|\mathbf A'(z)\|_2.
\]
Applying this to $\mathbf E$ gives the stated upper bound. For any two square matrices $\mathbf A,\mathbf B$ and any unit vector $\mathbf x$, $\|\mathbf A\mathbf x\|_2\ge\|\mathbf B\mathbf x\|_2-\|\mathbf A-\mathbf B\|_2$. Taking the infimum over $\mathbf x$ and then exchanging $\mathbf A,\mathbf B$ proves
\[
  |\sigma_{\min}(\mathbf A)-\sigma_{\min}(\mathbf B)|
      \le\|\mathbf A-\mathbf B\|_2.
\]
Use this inequality with $\mathbf D_0(\omega)$ and $\mathbf D_0(z_j)$ to obtain the lower bound, then take extrema over all arcs. Since $\overline\delta_\Gamma\ge0$, a successful strict test also implies $\underline\mu_\Gamma>0$.
\end{proof}

If the numerically evaluated matrices $\widehat{\mathbf E}_j$ and $\widehat{\mathbf D}_{0,j}$ at $z_j$ have evaluation error bounds $\epsilon_{E,j}$ and $\epsilon_{0,j}$, we can take
\begin{equation}
  u_{E,j}=\widehat n_j+\eta_{n,j}+\epsilon_{E,j},
  \qquad
  l_{0,j}=\widehat s_j-\eta_{s,j}-\epsilon_{0,j},
  \label{eq:app-floating-nodes}
\end{equation}
where $\widehat n_j+\eta_{n,j}\ge\|\widehat{\mathbf E}_j\|_2$ and $\widehat s_j-\eta_{s,j}\le\sigma_{\min}(\widehat{\mathbf D}_{0,j})$ bound the norm and minimum singular value. The scalar bounds include errors in matrix evaluation and subsequent arithmetic. Note that $a_j$ in Eq.~\eqref{eq:app-arc-radius} can be replaced by the chord length $|\omega-z_j|$ only if the derivative bounds in Eq.~\eqref{eq:boundary-derivative-bounds} hold on the line segment between $z_j$ and $\omega$.

\begin{lemma}[Derivative bound for a rational matrix function]
\label{lem:app-derivative}
Let
\[
  \mathbf A(\omega)=\sum_{k=0}^{d}\mathbf A_k\omega^k
     +\sum_g\sum_{\ell=1}^{\nu_g}
       \frac{\mathbf T_{g,\ell}}{(\omega-\xi_g)^\ell}.
\]
If $R_j\ge\max_{\omega\in\Gamma_j}|\omega|$ and $d_{g,j}\le\min_{\omega\in\Gamma_j}|\omega-\xi_g|$ with $d_{g,j}>0$, then
\begin{equation}
  \sup_{\omega\in\Gamma_j}\|\mathbf A'(\omega)\|_2
    \le\sum_{k=1}^{d}k\|\mathbf A_k\|_2R_j^{k-1}
       +\sum_g\sum_{\ell=1}^{\nu_g}
          \frac{\ell\|\mathbf T_{g,\ell}\|_2}{d_{g,j}^{\ell+1}}.
  \label{eq:app-rational-derivative}
\end{equation}
\end{lemma}

\begin{proof}
Differentiate each polynomial and pole term, apply the triangle inequality, and use the stated magnitude and distance bounds.
\end{proof}

For $\mathbf E=\mathbf D_1-\mathbf D_0$, if both matrices are rational, one can either combine their terms at shared poles before applying Lemma~\ref{lem:app-derivative} or bound $\|\mathbf D_1'\|_2$ and $\|\mathbf D_0'\|_2$ separately.

\subsection{The reference dispersion matrix and the total error bound}
\label{app:error-budget}

To bound $\|\mathbf D_\star-\mathbf D_0\|_2$ on $\Gamma$, consider a sequence of dispersion matrices $\mathbf T_0=\mathbf D_0,\mathbf T_1,\ldots,\mathbf T_K=\mathbf D_\star$ analytic on the same neighborhood of $\overline U$, where each step changes one approximation (such as the SVD truncation, the pole order $J$, the harmonic cutoff $N$, or the equilibrium distribution fit) under the same field variables, frequency normalization, and matrix scaling. If
\begin{equation}
  \sup_{\omega\in\Gamma}\|\mathbf T_j(\omega)-\mathbf T_{j-1}(\omega)\|_2
     \le\delta_j,
  \label{eq:app-stage-bounds}
\end{equation}
then telescoping and the triangle inequality give
\begin{equation}
  \sup_{\omega\in\Gamma}\|\mathbf D_\star(\omega)-\mathbf D_0(\omega)\|_2
  \le\sum_{j=1}^K\delta_j.
  \label{eq:app-telescoping}
\end{equation}
Numerical evaluation errors enter the bounds for these exact matrix differences and are counted only once. In particular, comparing two finite truncations (such as two pole orders $J_1,J_2$ or harmonic cutoffs $N_1,N_2$) bounds only $\|\mathbf T_j-\mathbf T_{j-1}\|_2$; bounding the distance to $\mathbf D_\star$ requires an additional bound on the truncation remainder.

\Needspace{10\baselineskip}
\subsection{Sensitivity of a simple root and sign of the growth rate}
\label{app:growth}

\begin{proposition}[Derivative of an isolated simple root]
\label{prop:app-sensitivity}
Suppose $\mathbf D_0$ and $\mathbf E$ are analytic near $\widehat\omega$ and $\widehat\omega$ is a simple root of $\det\mathbf D_0(\omega)=0$. Let $\mathbf v$ and $\mathbf w$ be nonzero right and left null vectors of $\mathbf D_0(\widehat\omega)$. For $\mathbf D_t=\mathbf D_0+t\mathbf E$, there is a unique analytic eigenvalue branch $\omega(t)$ near $t=0$ with $\omega(0)=\widehat\omega$, and
\begin{equation}
  \omega'(0)=-\frac{\mathbf w^*\mathbf E(\widehat\omega)\mathbf v}
                        {\mathbf w^*\mathbf D_0'(\widehat\omega)\mathbf v},
  \qquad \mathbf w^*\mathbf D_0'(\widehat\omega)\mathbf v\ne0.
  \label{eq:app-root-derivative}
\end{equation}
\end{proposition}

\begin{proof}
Set $f(t,\omega)=\det[\mathbf D_0(\omega)+t\mathbf E(\omega)]$. It is holomorphic in both variables, $f(0,\widehat\omega)=0$, and $\partial_\omega f(0,\widehat\omega)\ne0$. The holomorphic implicit function theorem gives the stated root branch with $\omega'(0)=-\partial_t f/\partial_\omega f$ at $(0,\widehat\omega)$.

Writing $\mathbf A=\mathbf D_0(\widehat\omega)$, simplicity implies $\operatorname{rank}\mathbf A=p-1$: if the rank were at most $p-2$, every cofactor would vanish and so would the determinant derivative. The adjugate is therefore nonzero and has the form $\operatorname{adj}(\mathbf A)=c\mathbf v\mathbf w^*$ for a nonzero scalar $c$, because its columns and rows lie in the one-dimensional right and left kernels. This also holds for $p=1$, with the usual scalar adjugate. The determinant differential formula, valid also at singular matrices, gives
\[
  \partial_t f=c\mathbf w^*\mathbf E(\widehat\omega)\mathbf v,
  \qquad
  \partial_\omega f=c\mathbf w^*\mathbf D_0'(\widehat\omega)\mathbf v.
\]
The latter is nonzero, and their ratio proves Eq.~\eqref{eq:app-root-derivative}.
\end{proof}

In particular,
\[
  |\omega'(0)|\le
  \frac{\|\mathbf w\|_2\|\mathbf v\|_2}
       {|\mathbf w^*\mathbf D_0'(\widehat\omega)\mathbf v|}
       \|\mathbf E(\widehat\omega)\|_2.
\]
This is the local sensitivity underlying the condition number of a simple eigenvalue of a nonlinear eigenvalue problem~\cite[Sec.~2.7, Theorem~2.20]{guttelTisseur2017nep}.

\begin{corollary}[Sign of the growth rate]
\label{cor:app-growth}
Let $\omega_1$ be a root of the reference dispersion matrix $\mathbf D_1$ and set $\gamma_1=\operatorname{Im}\omega_1$. If $|\omega_1-\widehat\omega|\le\rho$, then
\begin{equation}
  \operatorname{Im}\widehat\omega-\rho
     \le\operatorname{Im}\omega_1
     \le\operatorname{Im}\widehat\omega+\rho.
  \label{eq:app-growth-interval}
\end{equation}
Under Eq.~\eqref{eq:plane-wave-convention}, $\operatorname{Im}\widehat\omega-\rho>0$ implies $\gamma_1>0$ (instability), and $\operatorname{Im}\widehat\omega+\rho<0$ implies $\gamma_1<0$ (damping). More generally, if a contour $\Gamma$ satisfying Theorem~\ref{thm:app-contour} lies entirely in the open upper or lower half-plane, then all enclosed roots of $\mathbf D_1$ have the corresponding sign of $\operatorname{Im}\omega_1$.
\end{corollary}

\begin{proof}
Equation~\eqref{eq:app-growth-interval} follows from $|\operatorname{Im}(\omega_1-\widehat\omega)|\le|\omega_1-\widehat\omega|\le\rho$. For a general contour $\Gamma$ contained in the open upper or lower half-plane, the interior domain $U$ lies in the same half-plane, and rescaling $\omega$ by any positive normalization constant preserves the sign of $\operatorname{Im}\omega$.
\end{proof}

\end{document}